\documentclass[aps,prc,onecolumn]{revtex4}
\usepackage{epsf,epsfig,amssymb,amsmath,mathtools}

\usepackage{color}

\newcommand\SKIP[1]{}
\newcommand{\be}{\begin{equation}}
\newcommand{\ee}{\end{equation}}
\newcommand{\bea}{\begin{eqnarray}}
\newcommand{\eea}{\end{eqnarray}}
\newcommand{\mybibitem}{\bibitem}

\newcommand{\ch}{{\rm ch}}
\newcommand{\sh}{{\rm sh}}

\renewcommand{\vec}[1]{{\bf #1}}

\newcommand{\gton}{\mathrel{\lower.9ex \hbox{$\stackrel{\displaystyle 
>}{\sim}$}}} 
\newcommand{\lton}{\mathrel{\lower.9ex \hbox{$\stackrel{\displaystyle 
<}{\sim}$}}}

\newcommand{\vp}{{\bf p}}

\begin{document}

\title{Testing and improving shear viscous phase space 
correction models}

\author{Mridula Damodaran}
\author{Denes Molnar}
\affiliation{Department of Physics and Astronomy, Purdue University, West Lafayette, IN 47907}
\affiliation{Wigner Research Center for Physics, Budapest, Hungary}
\author{Gergely G\'abor Barnaf\"oldi}
\author{D\'aniel Ber\'enyi}
\author{M\'at\'e Ferenc Nagy-Egri}
\affiliation{Wigner Research Center for Physics, Budapest, Hungary}

\date{\today}

\begin{abstract}
Comparison of hydrodynamic calculations with experimental data inevitably requires
a model for converting the fluid to particles.
In this work, nonlinear $2\to 2$ kinetic theory is used to assess the
overall accuracy of various shear viscous fluid-to-particle conversion
models, such as the quadratic Grad corrections, 
the Strickland-Romatschke (SR) ansatz, self-consistent shear corrections
from linearized kinetic theory, and the correction from the relaxation time approach.
We test how well the conversion models can reconstruct, using solely the hydrodynamic fields computed from the transport, the phase space density
for a massless one-component gas undergoing a 0+1D longitudinal boost-invariant
expansion with approximately constant specific shear viscosity in the range 
$0.03 \lton \eta/s \lton 0.2$.

In general we find that at early times the SR form is the most accurate,
whereas at late times or for small $\eta/s\sim 0.05$ the self-consistent corrections
from kinetic theory perform the best. 
In addition, we show that the reconstruction accuracy of additive shear viscous
$f = f_{\rm eq} + \delta f$ models dramatically improves if one ensures,
through ``exponentiation'', that $f$ is always positive.
We also illustrate how even more accurate viscous $\delta f$ models can be
constructed if one includes information about the past evolution of the system 
via the first time derivative of hydrodynamic fields.
Such time derivatives 
are readily available in hydrodynamic simulations, though usually not included
in the output.
\end{abstract}

\maketitle

\section{Introduction}

Relativistic viscous hydrodynamics is 
probably the most popular framework to model ultrarelativistic heavy-ion
collisions (for reviews, see, e.g., 
Refs.~\cite{Huovinen:2006jp,Gale:2013da,Heinz:2013th}).
Its strong appeal is that it describes the evolution 
directly utilizing bulk matter properties
such as the equation of state and the shear and bulk viscosities. However, 
at least for heavy-ion physics applications, the framework
is incomplete. First, the early evolution is far from local thermal equilibrium; 
therefore, a model of initial conditions is needed. 
Second,
comparison with experimental data inevitably requires
a model for converting the fluid to particles. This so-called ``particlization'' 
\cite{Huovinen:2012is,Molnar:2011kx,MolnarWolff},
which is our focus here, 
is necessary irrespectively of whether
one compares hydrodynamics directly with experiments, or combines hydrodynamics with
subsequent hadronic transport (``hybrid'' approach) 
\cite{Song:2010aq,Hirano:2012kj,Ryu:2012at}.

For fluids in perfect local equilibrium, i.e., 
{\em ideal} fluids, the conversion to particles is straightforward, at least as long as
one models the fluid, at the conversion point, as an ideal gas of hadrons. 
Then, the phase space densities of the various hadron species are uniquely determined by the 
hydrodynamic fields. In contrast, {\em viscous} fluids are in general out of local equilibrium,
which makes the conversion ambiguous because an infinite class of phase space densities 
reproduces the same hydrodynamic fields\cite{MolnarWolff}
(this is so even for a one-component gas). 
Several different particlization models are in use based on
Grad's quadratic form\cite{Israel:1979wp,Grad1949}, 
the Romatschke-Strickland ansatz\cite{Romatschke:2003ms,Florkowski:2014bba}, the 
linearized Boltzmann transport equation\cite{MolnarWolff,AMYtrcoeffs,Dusling:2009df},
or kinetic theory in the relaxation time 
approach\cite{Florkowski:2014sfa,Dusling:2009df};
all of which originate from kinetic theory.

It is not clear {\em a priori} which particlization schemes, if any, 
are accurate. 
In this paper we investigate how well different
approaches can reproduce {\em shear} corrections generated by 
fully nonlinear $2\to 2$ transport theory.
The study here shares similarities with an earlier work~\cite{Huovinen:2008te} by one of us,
in which transport solutions obtained with Molnar's Parton Cascade (MPC) \cite{MPC} 
for a one-component gas in a 0+1D 
Bjorken scenario were used to test formulations 
of viscous hydrodynamics.
However, instead of studying how well hydrodynamic models follow 
the evolution of the energy-momentum tensor in the transport,
here we investigate how well
particlization models reproduce the actual phase space distribution in 
the transport
from the exact hydrodynamic fields corresponding to the transport solution.

In addition, we present two improvements to current shear viscous $\delta f$ models. 
First, we cure
the unphysical negative contributions in additive $f = f_{eq} + \delta f$ models,
and show that this leads to a dramatic improvement in accuracy. Moreover,
we illustrate how knowledge of the first time derivative of hydrodynamic fields,
which is readily available in hydrodynamic calculations, can be used to intelligently
switch between viscous correction models to apply them closer to their respective 
regions of validity.

The structure of the paper is as follows. A brief introduction to the
general problem in Sec.~\ref{Sc:df} is followed by a discussion of four
different shear viscous $\delta f$ models 
in Sec.~\ref{Sc:models}. Next, 
Sec.~\ref{Sc:transport} outlines 
covariant transport theory and the
MPC/Grid numerical transport solver,
and Sec.~\ref{Sc:tests} describes how the $\delta f$ models
are tested against kinetic theory. 
Finally, Sec.~\ref{Sc:results} presents
the results on the overall accuracy of the four shear viscous $\delta f$ models, 
as well as, novel improvements to current $\delta f$ models.

\section{Ambiguity in choosing viscous phase space corrections}
\label{Sc:df}

We briefly review here the challenge posed by the ambiguity in constructing
phase space distributions from hydrodynamics. Only the case of
a one-component system will be discussed 
(see, e.g., Ref.~\cite{MolnarWolff} for multi-component mixtures). 
For a noninteracting gas, the hydrodynamic fields, namely, the energy-momentum 
tensor
and the number current%
\footnote{
The number current is included
because, for the $2\to 2$ transport considered here, 
it is conserved.}
are directly
given by the (on-shell) phase space density $f \equiv dN/d^3 x d^3p$ as
\begin{equation}
T^{\mu\nu}(x) \equiv \int \frac{d^3p}{E} p^\mu p^\nu f(x, \vec{p})\ ,
\quad\quad\quad\quad 
N^\mu(x) \equiv \int \frac{d^3p}{E} p^\mu f(x, \vec{p}) \ .
\label{T_N}
\end{equation}
In local equilibrium%
\footnote{We consider Boltzmann statistics throughout, however, extension
to the Fermi/Bose case is straightforward.}
\begin{equation}
f^{\rm eq}(x, \vec{p}) 
= \frac{g}{(2\pi)^3} \exp \left[\frac{\mu(x) - p^\alpha u_\alpha(x)}{T(x)}\right] \ ,
\label{feq}
\end{equation}
which reproduces the fields in ideal (nonviscous) hydrodynamics:
\be
T^{\mu\nu}_{\rm eq} = (e + p) u^\mu u^\nu - p\, g^{\mu\nu} \ ,
\qquad\qquad
N^\mu_{\rm eq} = n u^\mu \ .
\label{hydro_eq}
\ee
Here, $g$ is the number of internal degrees of freedom, 
$e$, $p$, $n$, $u$, $T$, and $\mu$ are the local energy density, 
equilibrium pressure, particle density, flow
velocity, temperature, and 
chemical potential, respectively,
and we dropped the space-time argument $x$ for brevity.
For ideal fluids the phase space density is uniquely determined by
the hydrodynamic fields because straightforward inversion of (\ref{hydro_eq}) 
gives
\be
n = \sqrt{N_{\rm eq}^\mu N_{\rm eq,\mu}}\ , 
\qquad\qquad
u^\mu = \frac{N_{\rm eq}^\mu}{n} \ ,
\qquad\qquad
e = u_\mu T_{\rm eq}^{\mu\nu} u_\nu \ ,
\ee
and through the equations of state $p(e,n)$, $T(e,n)$, and $\mu(e,n)$ 
these give the local temperature and chemical potential.
Specifically, for the massless Boltzmann particles considered later here,
\be
p = \frac{e}{3} \ , 
\quad
T = \frac{p}{n} = \frac{e}{3n} \ ,
\quad
\mu = T \ln \frac{n}{n_{\rm eq}(T)} = T \ln \frac{27\pi^2 n^4}{g e^3} \ 
\ee
($n_{{\rm eq}} = g T^3/\pi^2$ is the thermal particle density at 
$\mu = 0$, i.e., in local 
thermal {\em and} chemical equilibrium). 

In the general out-of-local-equilibrium case, one can
split the phase space density into a local equilibrium piece and
a dissipative correction as $f(x,\vp) \equiv f_{\rm eq}(x,\vp) + \delta f(x,\vp)$
\footnote{
The decomposition depends on the choice of the fluid flow velocity field $u^\mu(x)$
that enters in $f_{\rm eq}$ (cf. (\ref{feq})).
Here we follow the Landau prescription that a local observer at $x$ 
comoving with the fluid should find no energy flow at $x$.
}.
Consequently, nonequilibrium corrections arise to the hydrodynamic fields as well:
\be
\delta T^{\mu\nu}(x) = \int \frac{d^3p}{E} p^\mu p^\nu \delta f(x, \vec{p}) \ , \qquad
\delta N^\mu(x) = \int \frac{d^3p}{E} p^\mu \delta f(x, \vec{p})  \ .
\label{delta_T_N}
\ee
Here $\delta T^{\mu\nu}$ contains shear stress and bulk pressure corrections,
whereas $\delta N^{\mu}$ describes particle diffusion.
While dissipative hydrodynamics solutions provide both the ideal part and the dissipative 
corrections
to all hydrodynamic fields, 
phase space densities are not available from hydrodynamics
because the  correction $\delta f$ is not known.
The challenge to particlization models is to invert (\ref{delta_T_N}) for 
$\delta f$. Inversion is only possible with additional theory input 
because an infinite class of different phase space distributions reproduces
the same hydrodynamic fields\cite{Wolff:2016vcm}.

\section{Models for shear viscous phase space corrections}
\label{Sc:models}

From here on we focus on phase space corrections due to shear only,
i.e., we take 
\be
\delta T^{\mu\nu}(x) = \pi^{\mu\nu}(x) , \quad \delta N^\mu(x) = 0 \ ,
\ee
where $\pi^{\mu\nu}(x)$ is the local shear stress. It is a symmetric,
traceless tensor that is purely spatial in the local rest (LR) 
frame of the fluid:
\be
u_\mu \pi^{\mu \nu } = 0 = \pi^{\nu\mu} u_\mu \ ,
\quad \pi^\mu_{\ \mu} = 0 \ .
\ee
(In the LR frame, $u_{LR}^\mu = (1, \vec 0)$.)
Below we review popular parametrizations for shear viscous phase space 
corrections.

\subsection{Grad ansatz}
\label{Sc:Grad}

A natural starting point 
for near-equilibrium phase space distributions
is an expansion in small gradients around local
equilibrium
\cite{Grad1949,Israel:1979wp,Huovinen:2008te,MolnarWolff} 
(also, Ch. VII of \cite{deGroot}):
\begin{equation}
f(x,\vec{p}) = f_{eq}(x,\vec{p})\left[1 + \phi(x,\vec{p})\right]
\qquad \text{with} \quad |\phi| \ll 1 \ , \quad 
|p^\mu \partial_\mu \phi| \ll |p^\mu \partial_\mu f_{eq}|/f_{eq} \ .
\label{BTEexp}
\end{equation}
The commonly used Grad ansatz then comes from a Taylor expansion of $\phi$ 
in powers of the momentum:
\begin{equation}
\phi (x, \vec{p}) = D^\mu(x) p_\mu + C^{\mu\nu}(x) p_\mu p_\nu + \mathcal{O}(p^3) \ .
\end{equation}
If there is no particle diffusion and no bulk pressure either%
\footnote{For the massless gas considered later in this work, bulk pressure
vanishes identically because the energy-momentum tensor (\ref{T_N}) is traceless.
}, then
$D^\mu = 0$,
which leads to
\begin{equation}
\phi_{Grad} (x, \vec{p}) = \frac{\pi^{\mu\nu}}{2(e + p)} \frac{p_\mu p_\nu}{T^2} = \frac{\pi^{\mu\nu}}{8p} \frac{p_\mu p_\nu}{T^2} \ ,
\label{eq:ISphi}
\end{equation}
where in the last step we specialized to a gas of massless particles 
($e = 3p$).
The $p_\mu p_\nu \pi^{\mu\nu}$ term introduces a characteristic quadratic momentum dependence.
The Grad ansatz also played a key role in the Israel-Stewart formulation of
 causal relativistic viscous hydrodynamics \cite{Israel:1979wp}.

\subsection{Strickland-Romatschke form (SR)}
\label{Sc:SR}

Another class of nonthermal phase space distributions has been proposed
based on stretching a spherically symmetric momentum 
distribution \cite{Romatschke:2003ms}:
\be
f(|\vp|) \to f\!\left(\sqrt{p_T^2+a^2 p_z^2}\right) \ ,
\qquad p_T \equiv \sqrt{p_x^2 + p_y^2} \ .
\ee
This form 
was originally motivated as a convenient way
to introduce transverse vs longitudinal momentum anisotropy via a
single 
parameter $a$. For transversely homogeneous, longitudinally
boost invariant dynamics with massless particles, the Strickland-Romatschke
(SR) ansatz reads
\be
f_{SR}(x,\vec{ p}) \equiv f_{eq} + \delta f_{SR} = N \exp\!\left(-\frac{1}{\Lambda}\sqrt{p_{LR,T}^2 + a^2 p_{LR,z}^2}\right) \ .
%
\ee
The advantage of the SR ansatz is that it is strictly positive everywhere, 
and for $a = 0$ it describes 
local thermal equilibrium, while for $a = \tau/\tau_0$ it gives back the 
free streaming evolution for massless
particles as long as the system started at $\tau = \tau_0$ from local thermal equilibrium. 
Here $\tau$ is the Bjorken proper time (cf. Sec.~\ref{Sc:setup}).

More complicated anisotropies can be accommodated via suitable generalizations of the ansatz \cite{Florkowski:2014bba,Tinti:2015xwa}.
The SR ansatz and its extensions are also employed 
in the formulation of anisotropic hydrodynamics\cite{Martinez:2010sc}.

\subsection{Self-consistent viscous corrections from linearized covariant transport}
\label{Sc:lin}

In contrast to ad-hoc parametrizations, 
self-consistent viscous corrections can be obtained
from relativistic kinetic theory\cite{MolnarWolff}.
These follow from the relativistic Boltzmann transport 
equation (BTE), if one linearizes in the departure from local
equilibrium, and studies the late-time asymptotic evolution
in the presence of flow gradients in the system.
We refer the Reader to 
Refs.~\cite{MolnarWolff,AMYtrcoeffs,Dusling:2009df}
for details of the procedure. For shear, one can show that
\be
\delta f_{lin} = \chi\!\left(\frac{p\cdot u}{T}\right) \frac{p_\mu p_\nu \sigma^{\mu\nu}}{T^3} f_{eq} \ ,
\ee
where the dimensionless function $\chi$ is the solution to a linear 
integral equation
\be
  p^\mu \nabla_\mu f_{\rm eq} = L[\delta f_{lin}] \ ,
\label{chiBTE}
\ee
 and the shear tensor 
\be
\sigma^{\mu\nu} \equiv \nabla^\mu u^\nu + \nabla^\nu u^\mu - \frac{2}{3} \Delta^{\mu\nu}(\partial \cdot u)
\ee 
characterizes shear deformation of the flow field.
The projector $\Delta^{\mu\nu} \equiv g^{\mu\nu} - u^\mu u^\nu$ 
and the gradient $\nabla^\mu \equiv \Delta^{\mu\nu} \partial_\nu$
orthogonal to the flow
ensure that $\sigma^{\mu\nu}$ is traceless and 
in the LR frame purely spatial.

In general, the integral equation must be solved for $\chi$
numerically.
Replacing the shear tensor with the shear stress tensor using the 
Navier-Stokes relation $\pi^{\mu\nu} = \eta \sigma^{\mu\nu}$ gives
\be
\delta f_{lin} = \chi\!\left(\frac{p\cdot u}{T}\right) \frac{p_\mu p_\nu \pi^{\mu\nu}}{\eta T^3} f_{eq} \ ,
\ee
which is the same as the quadratic Grad form but generalized to 
a self-consistently 
determined function of momentum.
For isotropic $2\to 2$ scattering with constant cross sections, 
the self-consistent viscous corrections
are well approximated\cite{MolnarWolff} by $\chi(x) \approx const / \sqrt{x}$,
i.e., $\delta f/f_{eq}$ has $\sim p^{3/2}$ momentum dependence.
Interestingly, the same 3/2 exponent arises from kinetic theory with forward-peaked
 $1\leftrightarrow 2$ perturbative QCD matrix elements\cite{Dusling:2009df}.

\subsection{Relaxation time approximation (RTA)}
\label{Sc:RTA}

Alternatively, one might treat the Boltzmann transport equation 
in the relaxation time approximation\cite{Florkowski:2014sfa,Dusling:2009df} 
(RTA):
\begin{equation}
p^\mu \partial_\mu f(x,\vec{p}) = - \frac{p\cdot u}{\tau_{eq}} \left(f(x,\vec{p}) - f_{eq}(x,\vec{p})\right) \ .
\label{RTAeq}
\end{equation}
At late times, the system relaxes to equilibrium and 
$\partial_t f \rightarrow 0$. In this regime, one can perform
a calculation analogous to that for the self-consistent viscous
corrections of Sec.~\ref{Sc:lin}
but with the simplified RTA kernel. For the RTA, (\ref{chiBTE}) reads
\be
p^\mu \nabla_\mu f_{\rm eq} = -\frac{p\cdot u}{\tau_{eq}} \delta f_{RTA} \ .
\ee
For pure shear deformation of the flow fields\cite{MolnarWolff}, 
\be
p^\mu \nabla_\mu f_{\rm eq} = -\frac{1}{2T}p_\mu p_\nu \sigma^{\mu\nu} f_{\rm eq} \ ,
\ee
which then leads to 
\be    
\delta f_{RTA} = \frac{\tau_{rel}}{2T (p\cdot u)} \sigma^{\mu\nu} p_\mu p_\nu f_{eq} \ .
\ee
Thus, if the relaxation time does not depend on momentum, then 
the shear correction from the RTA is
linear in momentum for massless particles.
(Other powers could be accommodated with appropriate
momentum dependent $\tau_{rel}$ \cite{Dusling:2009df}.)

\subsection{Power law generalization of Grad ansatz (PG)}
\label{Sc:powerlaw}

A simple generalization of the Grad shear correction (\ref{eq:ISphi})
is to include arbitrary $p^\alpha$ power ($\alpha > 0$) of 
the momentum via
\begin{equation}
\phi_{PG} (x, \vec{p}) = C(\alpha) \left(\frac{p\cdot u}{T}\right)^{\alpha-2}\; \frac{\pi^{\mu\nu}}{8p} \frac{p_\mu p_\nu}{T^2} \ , \qquad C(\alpha) = \frac{5!}{(3+\alpha)!} \ .
\label{genGrad}
\end{equation}
The coefficient $C(\alpha)$ is set by the requirement 
that $\phi$ gives back the correct shear stress through (\ref{delta_T_N}),
and we have used that the particles are massless (this form
was also discussed in \cite{Dusling:2009df}).

Setting $\alpha = 2$ gives back the quadratic Grad ansatz of Sec.~\ref{Sc:Grad}. 
On the other hand, as we saw in Sec.~\ref{Sc:lin},
shear viscous $\delta f$ calculations
based on linearized kinetic theory
are well approximated by a significantly smaller $\alpha \approx 3/2$. 
Shear viscous corrections based on the 
relaxation time approximation (Sec.~\ref{Sc:RTA})
correspond to a yet smaller $\alpha = 1$ exponent 
(if the relaxation time is constant).
Thus, the power-law Grad form can conveniently capture three rather different models.
{\em In what follows we are going to concentrate on four different shear viscous 
$\delta f$ models:
the power-law Grad ansatz with exponents
$\alpha = 1$ (RTA), $3/2$ (full BTE), and $2$ (Grad), and the Strickland-Romatschke form.}

A shortcoming of both the Grad ansatz and its power law
extension is that at high momenta the phase space 
distribution $f$ can become negative. The shear stress tensor 
is traceless, so it has positive and negative
eigenvalues. Therefore, $p_\mu p_\nu \pi^{\mu\nu}$ is not bounded from below. 
A possible
remedy through ``exponentiation'' is discussed in Sec.~\ref{Sc:RMS}.

\section{Covariant transport theory}
\label{Sc:transport}

Let us now turn to discussing covariant transport theory that
will be used to test the accuracy of shear viscous $\delta f$ models.

\subsection{Boltzmann transport equation}

Consider on-shell covariant transport theory for a system with
$2\to 2$ interactions as, e.g., in Refs. \cite{Molnar:2000jh,Huovinen:2008te} 
(for a discussion of the multicomponent case, see \cite{MolnarWolff}). 
The evolution of
the phase space density is given by the nonlinear Boltzmann transport
equation
\be
p^\mu \partial_\mu f(x,\vp) = S(x,\vp) +
C[f](x, \vp) \ ,
\label{BTE}
\ee
where the source term $S$ encodes the initial conditions,
and the two-body collision term is
\be
C[f](x,\vp_1) 
\equiv \int\limits_2 \!\!\!\!\int\limits_3 \!\!\!\!\int\limits_4
\left(f_{3} f_{4} - f_{1} f_{2}\right)
\, \bar W_{12\to 34}  \, \delta^4(12 - 34)
\label{Cijkl}
\ee
with shorthands
$\int\limits_a \equiv \int d^3p_a / (2 E_a)$, 
$f_{a} \equiv f(x,\vp_a)$, and
$\delta^4(ab - cd) \equiv \delta^4(p_a + p_b - p_c - p_d)$.
The transition probability $\bar W_{12\to 34}$ for $2\to 2$ scattering
with momenta 
$p_1 + p_2 \to p_3 + p_4$ is given by the differential cross section as
$\bar W = 4 s d\sigma / d\Omega_{cm}$, where $s \equiv (p_1 + p_2)^2$ 
is the usual Mandelstam variable, and the
solid angle is taken in the c.m. frame of the microscopic two-body collision.

\subsection{Numerical solutions via the MPC/Grid algorithm}

To compute transport solutions numerically, we utilize a new parallelized version
of the MPC/Grid solver
from Molnar's Parton Cascade (MPC) \cite{MPC}. Details of the algorithm will be published
elsewhere, so here we only focus on the key ingredients.
Similarly to the Boltzmann Approach to Multi-Parton Scattering code
(BAMPS) \cite{Xu:2004mz},
MPC/Grid uses test particles on a three-dimensional spatial grid,
which undergo random scatterings. Between scatterings, the test particles move 
on straight lines.
In each time step $\Delta t$, in each cell, 
pairs of test particles are tested for $2\to 2$ scattering with 
probability $P_{2\to 2} = \sigma v_{rel} \Delta t /V_{cell}$, where $\sigma$ 
is the total cross section, $v_{rel}\equiv \sqrt{(p_1 \cdot p_2) - m_1^2 m_2^2}/E_1E_2$ is the relative velocity of the pair,
and $V_{cell}$ is the volume of the cell. 
If a scattering
occurs, outgoing momenta are generated according to the differential cross section.
MPC/Grid can also take $2\leftrightarrow 3$ scatterings into account but 
those were not turned on in this study.

Formally, the grid algorithm obtains the correct transport solution
in the limit of small time steps, small cell sizes, and large number of test particles. This gives extra flexibility compared to cascade algorithms 
based on scattering at closest approach.
In the cascade approach, 
numerical artifacts can only be reduced via increasing the number of test 
particles through particle subdivision\cite{Zhang:1998tj,Molnar:2000jh} 
$N_{test} \to \ell N_{test}$. 
Therefore, in the cascade,
the effective range of nonlocal interactions decreases as 
$1/\sqrt{\ell}$ jointly
in all
three principle directions.
In contrast, the grid approach controls
the effective range of nonlocality independently in 
each principle direction through the cell sizes in those directions,
which enables much faster calculations in situations with high symmetry.
For example, in the 0+1D
(transversely homogeneous and static) Bjorken scenario of Sec.~\ref{Sc:tests}, 
the cell size only needs to be small in the rapidity $\eta$ direction, but in the 
transverse directions it can stay arbitrarily large. Reduction
of the effective range by a factor $\lambda$ requires 
${\cal O}(\lambda^4)$ times longer computation with the cascade method,
but not worse than ${\cal O}(\lambda^3)$ with the grid approach
(in fact, only ${\cal O}(\lambda)$ if $N_{test}$ is sufficiently large).

\section{Testing $\boldsymbol{\delta f}$ models}
\label{Sc:tests}

Let us turn to the specifics of the transport calculations
and comparisons used to gauge the accuracy of viscous $\delta f$ models.

\subsection{Calculation setup}
\label{Sc:setup}

To study the accuracy of shear viscous $\delta f$ models, we consider a 
system of massless particles
evolving in a 0+1D Bjorken scenario. 
Transversely, the system is static, homogeneous, and
rotation invariant (in the actual calculation, periodic boundary conditions are imposed). 
Longitudinally, the system undergoes
boost-invariant%
\footnote{
By longitudinal boost invariance we mean that the state of the system
at each point in spacetime with $t>0$,  $z \ne 0$ 
can be obtained from the state on the $z = 0$ sheet via Lorentz
boost along the $z$ direction.
}
 Bjorken expansion, and we also impose $z\to -z$ reflection symmetry. 
This scenario has been extensively studied, e.g.,
in \cite{Huovinen:2008te}, so we only remind the Reader of a few important features here. 

A convenient set of coordinates is provided by 
the Bjorken proper time $\tau$, coordinate rapidity 
$\eta$, momentum rapidity $y$, and transverse momentum $p_T$, defined via
\be
\tau \equiv \sqrt{t^2 - z^2} \ , \quad \eta \equiv \frac{1}{2}\ln \frac{t+z}{t-z} \ ,
\quad y \equiv \frac{1}{2} \ln \frac{E+p_z}{E - p_z} \ ,
\quad p_T \equiv \sqrt{p_x^2 + p_y^2} \ .
\ee
Due to the high degree of symmetry, the phase space density only depends on $p_T$, $\xi \equiv \eta - y$, and $\tau$.
Since $f(p_T,\xi,\tau) = f(p_T,-\xi,\tau)$, it is often convenient to consider
that the $\xi$ dependence is through $\ch\, \xi$.
The local flow velocity and particle density of the system are
\be
u^\mu = (\ch\, \eta, \vec 0_T, \sh\,\eta) \ , \quad
n(\tau) = \frac{\tau_0}{\tau} n_0 \ ,
\ee
where $n_0 \equiv n(\tau_0)$ and
\be
n(\tau)  = 4\pi\int_0^\infty dp_T\, p_T^2 \int_0^\infty d\xi\, \ch \xi\, f(\xi,p_T,\tau) \ .
\label{n_with_f}
\ee
The energy momentum tensor is diagonal in the LR frame,
and because bulk viscosity and the bulk viscous 
pressure vanish for massless particles, it reads
\be
T^{\mu\nu}_{LR} = T^{\mu\nu}(\eta = 0) = diag\!\left(e, p-\frac{\pi_L}{2}, p - \frac{\pi_L}{2}, p + \pi_L\right) \quad {\rm with}
\quad p = \frac{e}{3} \ .
\ee
There are only two remaining independent hydrodynamic variables, 
the local comoving energy density and longitudinal shear stress, which are given by
\bea
e(\tau) &=& \int \frac{d^3p}{E} (p\cdot u)^2\, f 
= 4\pi \int_0^\infty dp_T\, p_T^3 \int_0^\infty d\xi \ch^2 \xi\, f(p_T,\xi,\tau) \ ,
\label{e_with_f}
\\
\pi_L(\tau) &\equiv& T^{zz}_{LR} - p = \int \frac{d^3p}{E} \left(p_{z,LR}^2 - \frac{p_{LR}^2}{3}\right) f = 4\pi \int_0^\infty dp_T\, p_T^3 
           \int_0^\infty d\xi \left(\sh^2 \xi - \frac{\ch^2 \xi}{3}\right) f(p_T,\xi,\tau) \ .
\label{piL_with_f}
\eea

It will be more convenient later to use equivalent expressions
in terms of the momentum
distribution of particles%
\footnote{
The momentum distribution of particles crossing
a three-dimensional hypersurface at spacetime coordinate $x$ is\cite{Cooper:1974mv}
$$
E\frac{dN}{d^3p} = p^\mu d\sigma_\mu(x)\, f(x,\vp) \ ,
$$
where $d\sigma^\mu$ is the Minkowski normal to the hypersurface at $x$.
For $\tau = const$ hypersurfaces, 
$p^\mu d\sigma_\mu = m_T\, \ch \xi\, \tau d^2 x_T d\eta$, where $m_T \equiv \sqrt{p_T^2 + m^2}$ and $\vec x_T$ are the transverse mass and position. 
Substitution
of  $d^3p/E \equiv d^2p_T dy$ with a switch 
to polar coordinates in the transverse plane leads to 
(\ref{dNdptdxi}) for massless particles.
}
\be
F(p_T,\xi,\tau) \equiv \frac{dN}{dp_T d\xi d\eta} 
= 2\pi A_T \tau p_T^2\, \ch \xi\, f(p_T,\xi,\tau) \ ,
\label{dNdptdxi}
\ee
where $A_T$ is the transverse area of the system. Then,
\bea
e(\tau) &=& \frac{2}{A_T\tau} \int_0^\infty dp_T\, p_T \int_0^\infty d\xi\, \ch \xi\, 
F(p_T, \xi,\tau) \ ,
\label{e_with_dN}
\\
\pi_L(\tau) &=& \frac{2}{A_T\tau} \int_0^\infty dp_T\, p_T 
           \int_0^\infty d\xi \left(\frac{\sh^2 \xi}{\ch \xi} - \frac{\ch \xi}{3}\right) F(p_T,\xi,\tau) \ ,
\label{piL_with_dN}
\\
n(\tau) &=& \frac{2}{A_T\tau} \int_0^\infty dp_T 
           \int_0^\infty d\xi\, F(p_T,\xi,\tau) \ .
\label{n_with_dN}
\eea
Due to longitudinal boost invariance, $F$ has no
separate $\eta$ dependence (it only depends on $\eta$ through $\xi$).

To mimic $\eta/s \approx const$ dynamics\cite{Huovinen:2008te}, 
isotropic $2\to 2$ cross
sections are used that grow with proper time
 as $\sigma(\tau) = (\tau/\tau_0)^{2/3} \sigma_0$.
We start the system at $\tau = \tau_0$ 
from local thermal equilibrium. 
If one uses dimensionless proper time $\tau/\tau_0$
and dimensionless momenta $p_T/T_0$,
then for massless particles 
the transport evolution depends only\cite{Molnar:2000jh} 
on the initial inverse Knudsen number
\be
K_0 \equiv \frac{\tau_0}{\lambda_{MFP}(\tau_0)} = n_0 \sigma_0 \tau_0 \ ,
\ee
which is the ratio of time scales for expansion and scattering.
The inverse Knudsen number is tightly related\cite{Huovinen:2008te} to the specific shear viscosity $\eta / s$,
which is the dimensionless ratio of the shear viscosity
and the entropy density.
Replacing $\sigma$ and $n$ with the shear viscosity $\eta\approx 1.2676T/\sigma$ and entropy density%
\footnote{For a massless gas in chemical equilibrium, $s = 4n$. Out of chemical equilibrium,
the factor of 4 shifts to $4 - \ln (n/n_{\rm eq})$, 
which is usually a modest correction only.}
$s \sim 4n$ one obtains
\be
K_0 \approx  \frac{T_0\tau_0}{5} \frac{s}{\eta} \ .
\ee
For A+A reactions at top Relativistic Heavy Ion Collider (RHIC) and Large Hadron Collider (LHC) energies, 
$T_0 \tau_0 \sim 1$ at particlization, 
so $\eta /s \sim 1/(5K_0)$. In what 
follows we will often refer to calculations by their approximate $\eta/s$ 
value because that is of high interest within
the community.

The transport evolution was computed for initial inverse Knudsen numbers in the
range $1 \le K_0 \le 6.47$ ($0.03 \lton \eta/s \lton 0.2$).
For each $K_0$ value,
high-statistics transport solutions corresponding to 
total 270 million test particles were calculated on the community clusters
at the Rosen Center for Advanced Computing (RCAC) at Purdue. To ensure approximately 
boost-invariant evolution, the systems were initialized with uniform 
coordinate rapidity distribution $dN/d\eta$ in a large initial 
coordinate rapidity window $|\eta| < 6$. Near the longitudinal 
$\eta \sim \pm 6$ edges, boost invariance
certainly does not apply. 
Therefore, only test particles close to midrapidity with $|\eta| < y_0$ are analyzed. 
Boost invariance
was monitored during the evolution through tracking $dN/d\eta$ vs $\eta$ and $\tau$.
To ensure that relative $dN/d\eta$ variations stay in all cases below
0.3\% in the midrapidity window, we set $y_0 = 2$,
which leaves about 90 million test particles for analysis at each value of $K_0$.

\begin{figure}[h]
\leavevmode
\epsfysize=6.5cm
\epsfbox{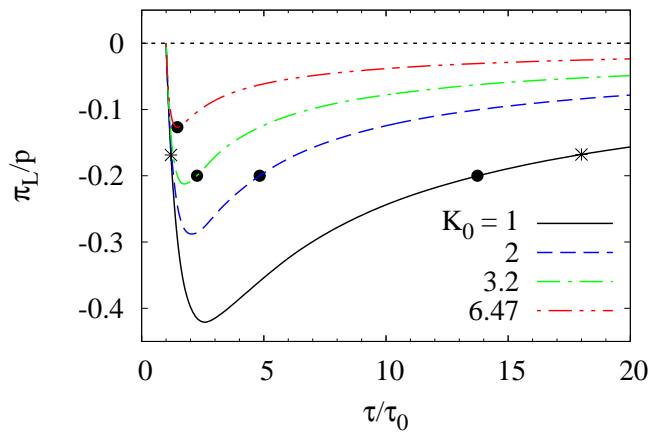}
\caption{Shear stress to equilibrium pressure ratio as a function of
rescaled proper time $\tau/\tau_0$ calculated with MPC/Grid 
for a massless system in 
a 0+1D Bjorken scenario with energy-independent,
isotropic $2\to 2$ cross section.
To mimic $\eta/s \approx const$ evolution, the cross section was
increasing with time proportionally to $\tau^{2/3}$.
Results for initial Knudsen numbers $K_0 = 1$ ($\eta/s\sim 0.2$, solid black), 
2 ($\eta/s \sim 0.1$, dashed blue), 3.2 ($\eta/s\sim 0.06$, 
dashed-dotted green), and 6.47 ($\eta/s\sim 0.03$, dashed-double-dotted red line) are shown. Stars
indicate the two time slices that are compared in Fig.~\ref{Fig:memory},
while filled circles show the switching times for the simple improved $\delta f$ 
model in Fig.~\ref{Fig:switch}.}
\label{Fig:Rpi}
\end{figure}

It is well known\cite{Huovinen:2008te} that for $\eta/s \approx const$, 
the rapid longitudinal Bjorken expansion first drives the system away from local equilibrium 
but later the system relaxes back to local equilibrium.
A convenient dimensionless parameter that measures the degree of departure from local 
equilibrium is the shear stress to pressure ratio
\be
 R_\pi \equiv \frac{\pi_L}{p}
\ee
(in local equilibrium $R_\pi = 0$). As shown in Fig.~\ref{Fig:Rpi}, 
during early times  $\pi_L$ (and thus $R_\pi$) becomes more and more negative 
but at late times $\pi_L$ vanishes asymptotically.
The larger $K_0$ is, the closer the system stays
to local thermal equilibrium. For $\eta/s \approx 0.2$ ($K_0 = 1$) shear
stress becomes fairly large in magnitude, reaching $|\pi_L| \approx 0.4p$.

\subsection{Fitting $\boldsymbol{\delta f}$ model parameters}
\label{Sc:fitting}

The primary output of the transport code is the momentum distribution 
(\ref{dNdptdxi}) of test particles crossing $\tau = const$ hypersurfaces,
given in terms of a list of test particles:
\be
F_{tr}(p_T, \xi,\tau) = \frac{1}{2y_0\ell} \sum_{i=1}^{N_{test}} \delta(p_{T,i} - p_T) \delta(\xi_i - \xi) \ .
\ee
It is straightforward to apply (\ref{e_with_dN}), (\ref{piL_with_dN}), and (\ref{n_with_dN}) to
calculate the comoving energy density, longitudinal shear stress, and number density
corresponding to the transport simulations ($1/\sqrt{N}$  relative 
fluctuations in these quantities
are about $0.1$\% with our statistics). Each simulation 
was then followed for each of the four $\delta f$ models
by an {\em inversion} of (\ref{e_with_f}) (\ref{piL_with_f}), and (\ref{n_with_f}) 
to match model parameters to 
the hydrodynamic fields at several proper times in the range 
$1 \le \tau/\tau_0 \le 20$
chosen
for analysis.

In Bjorken coordinates, the power-law Grad from and the Strickland-Romatschke ansatz read
\bea
f_{PG} &=& N \left[1 + \frac{15 R_\pi}{(3+\alpha)!} \left(\frac{p_T}{T}\right)^{\alpha}\,\ch^{\alpha-2}\xi\, \left(\sh^2\xi - \frac{1}{2}\right)\right] e^{-p_T \ch \xi / T} \ ,
\\
f_{SR} &=& N \exp\!\left(-\frac{p_T}{\Lambda} \sqrt{1+a^2 \sh^2 \xi}\right) \ .
\eea
Thus, all $\delta f$ models considered here have three parameters: 
an overall normalization factor $N$,
a momentum scale ($T$ or $\Lambda$), and a dimensionless pressure anisotropy
parameter
($R_\pi$ or $a$). It is, therefore, convenient to first match the anisotropy parameter
to reproduce $\pi_L/e$, then set the momentum scale from the effective temperature
\be
T_{eff} \equiv \frac{e}{3n} \ ,
\ee
and finally get the normalization from $n$. This way, each step
can be accomplished with root finding in only one dimension%
\footnote{
If in doubt, recast (\ref{e_with_dN}), (\ref{piL_with_dN}) and (\ref{n_with_dN})
in terms of dimensionless $q \equiv p_T/T$ for the power-law Grad form to
find
$$
e \propto NT^4 \ , \quad
\pi_L \propto NT^4 \ , \quad
n \propto NT^3 \ ,
$$
where the dimensionless proportionality constants only depend on $R_\pi$.
Analogous results follow for the SR ansatz but with $\Lambda$ instead of $T$,
and $a$ instead of $R_\pi$.
}
. The necessary 
$p_T$ and $\xi$ integrals were performed numerically.

\subsection{Quantifying model accuracy}
\label{Sc:RMS}

Our primary goal is to assess the accuracy of the four $\delta f$ models of
Sec.~\ref{Sc:models},
namely, how well they reconstruct the phase space density $f$, or equivalently,
the momentum distribution $F$. The most differential comparison would be to
study the relative reconstruction error in the $p_T-\xi$ plane 
as a function of the proper time $\tau$, given
by the ratio of the reconstructed distribution $F_{rec}$ to the actual
distribution $F_{tr}$ from the transport as
\be
 \varepsilon(p_T,\xi,\tau) \equiv \frac{F_{rec}(p_T,\xi,\tau)}{F_{tr}(p_T,\xi,\tau)} - 1  \ .
\label{epsFull}
\ee
It is more useful to take $F_{rec} / F_{tr}$ and not $F_{tr}/F_{rec}$
because that avoids poles if $F_{rec}$ goes negative.
Numerically, the transport distribution is represented by test particles, 
therefore it is more meaningful to compare two-dimensional histograms (integrated counts
over small $\Delta p_T \times \Delta\xi$ regions):
\be
\varepsilon_{ij}(\tau) \equiv \frac{\int_{p_{T,i-1}}^{p_{T,i}} dp_T
                                   \int_{\xi_{j-1}}^{\xi_j} d\xi\, F_{rec}(p_T,\xi,\tau)}{\Delta N_{tr}(p_{T,i-1} \le p_T < p_{T,i}, \xi_{j-1} \le \xi < \xi_j)} - 1\ ,
\qquad p_{T,i} \equiv i \Delta p_T \ , \quad \xi_j \equiv j \Delta \xi 
\quad (i=1,\dots,\ j=1,\dots)\ .
\label{epsDiscrete}
\ee 
However, this still gives too much information to present in a journal article.

Therefore, we compare here the root-mean-squared (RMS) 
relative error across all $p_T-\xi$ bins
for the various $\delta f$ models as a function of $\tau$:
\be
\varepsilon_{RMS}(\tau) \equiv \sqrt{\frac{1}{N_{bins}}\sum_{ij} \varepsilon_{ij}^2(\tau) } \ ,
\label{epsRMS}
\ee
which characterizes the overall accuracy in terms of a single number. 
This is, purposefully, quite 
different from the popular $\chi^2$ quantity used in data versus model 
comparisons because the terms in (\ref{epsRMS}) are not divided by a 
statistical 
error estimate for the measured counts. 
In our case, all reconstruction models are only approximate (strictly speaking, incorrect),
and we want to determine their overall error in $p_T-\xi$ space.
Therefore, as long as all counts in the histogram are measured with reasonable accuracy,
there is no inherent value in giving larger weight to bins with
very accurate counts.
With $\chi^2$ statistics, one would find for each model 
that $\chi^2$ grows with the size of the statistical sample, indicating that statistically
the models are less and less likely to be correct. In contrast, 
the root-mean-squared error converges with increasing sample size.

We evaluate $\varepsilon_{ij}$ within bins of size
$\Delta p_T = 0.16 T_0$ and $ \Delta \xi \approx 0.1$, 
across the rectangular region $p_T \le 12 T_{eff}$, $|\xi| \le 4$.
To limit statistical fluctuations, 
only bins with at least 200 counts are included in the average.
The two-dimensional integrals needed in (\ref{epsDiscrete}) 
were calculated via nesting one-dimensional
adaptive Gauss-Kronrod quadrature routines from the GNU Scientific Library (GSL) 
\cite{GSL},
and using an OpenCL implementation of 
two-dimensional Simpson quadrature
on graphics processor units (GPUs) at the GPU Laboratory of the Wigner Research
Center for Physics (Budapest, Hungary).

\section{Results and improved $\boldsymbol{\delta f}$ models}
\label{Sc:results}

In this section we quantify the accuracy of the four shear viscous $\delta f$ models
discussed in Sec.~\ref{Sc:models}, and also discuss novel improvements to these models.

\subsection{Reconstruction error and exponentiation}
\label{Sc:exp}

To characterize the accuracy of the various shear viscous $\delta f$ models
against fully nonlinear Bjorken evolution in 0+1D, one can use 
the overall RMS error $\varepsilon(\tau)$ introduced in Sec.~\ref{Sc:tests}.
Generally we find that at early times, errors are smallest for the SR ansatz 
(cf. Sec.~\ref{Sc:SR}), whereas
at late times or for large $K_0$ the self-consistent shear corrections are the 
the most accurate
(cf. Sec.~\ref{Sc:lin}), which are approximated here using the power-law Grad
form with exponent $\alpha = 3/2$. 
But one issue that plagues the accuracy of phase space
corrections linear in $\pi^{\mu\nu}$ is negativity. 
For 0+1D Bjorken
evolution, the correction for the power-law Grad form is proportional
to $R_\pi (\sh^2 \xi - 1/2)$,
which becomes negative for $|\xi| \gton 0.66$ because 
in our case $R_\pi < 0$ (cf. Fig.~\ref{Fig:Rpi}). 
Thus, at sufficiently high $p_T$, 
the total $f_{eq} + \delta f$ phase space density becomes negative,
which is unphysical.
The $p_T$ threshold where $f$ turns negative drops with increasing $|\xi|$.

We propose to cure the negative contributions via a simple exponentiation,
interpreting $1 + \phi$ in (\ref{BTEexp}) as the leading term of $e^\phi$ near $\phi = 0$.
This immediately renders all phase space densities positive, however,
it does not ensure that the integrals for the hydrodynamic fields
(\ref{n_with_f}), (\ref{e_with_f}), and (\ref{piL_with_f}) converge (e.g.,
for the Grad ansatz the exponent is then dominated by the quadratic terms, 
which are not positive definite). Therefore, we force the viscous
correction in the exponent to be bounded%
\footnote{
Alternatively, one could divide $\phi$ in the exponent by a momentum dependent
function that at large momenta fixes the asymptotic behavior whereas
at low momenta stays approximately constant.
}
via
\be
1 + \phi \to e^\phi \to e^{\tanh \phi} \ .
\label{expPhi}
\ee
For small $\phi$, this still reduces to $1 + \phi + {\cal O}(\phi^2)$.
Equation (\ref{expPhi}) is not the only possible solution; for example,
\be
1 + \phi \to \exp\!\left(\beta \tanh{\frac{\phi}{\beta}}\right)
\ee
accomplishes the same for any $\beta \ne 0$, 
and one can even take momentum dependent $\beta(p)$ here
(with some restrictions imposed by integrability).
In what follows, we refrain from fine-tuning and set $\beta = 1$.

\begin{figure}[h]
\leavevmode
\epsfysize=6.5cm
\epsfbox{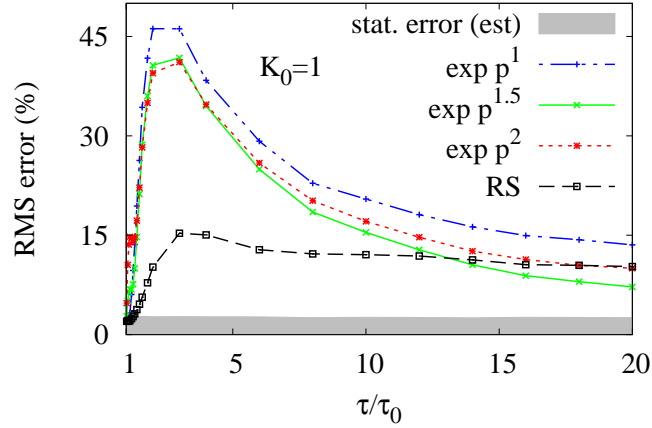}
\caption{Root-mean-squared phase space density 
reconstruction error as a function of
normalized proper time $\tau/\tau_0$ 
for a massless system undergoing  
a 0+1D Bjorken expansion with $\eta/s \sim 0.2$ 
(initial inverse Knudsen number $K_0 = 1$). Errors for the
Strickland-Romatschke ansatz (black dashed line) are compared
to those for the exponentiated power-law Grad ansatz with exponents
corresponding to the RTA model ($\alpha = 1$, dashed-dotted blue line), 
self-consistent $\delta f$ from 
Boltzmann transport ($\alpha = 3/2$, solid green line),
and the Grad ansatz ($\alpha = 2$, dotted red line). The shaded
gray band is the estimated statistical error in the analysis due to finite
test particle number.}
\label{Fig:RMS_K1}
\end{figure}

Figure~\ref{Fig:RMS_K1} shows the evolution of the overall RMS error as a function
of normalized proper time $\tau / \tau_0$ for initial inverse Knudsen number
$K_0 = 1$ ($\eta/s \sim 0.2$). At such high viscosity, the SR model 
(black dashed line)
has the smallest error for a wide range of times $\tau \lton 13\tau_0$. 
At $\tau = \tau_0$ it is completely accurate because we chose to start the evolution 
from local thermal equilibrium.
Moreover, at early times, errors can only accumulate quadratically with $\tau$ because 
to first order in time the transport evolution starts with {\em free 
streaming}\cite{Huovinen:2008te} (the collision term vanishes in local equilibrium),
which the SR ansatz can match exactly.
By $\tau \approx 3\tau_0$, however,
the RMS error reaches $15$\%, and it stays roughly flat in time,
decreasing only very slowly. Thus, overall, the SR ansatz is 10-15\% accurate
for $K_0 = 1$.

Figure~\ref{Fig:RMS_K1} also shows results for the power-law
Grad form with three exponents $\alpha = 1$
for the RTA model (dash-dotted blue line), 
3/2 for self-consistent shear corrections
from Boltzmann transport (solid green line),
and 2 for the Grad ansatz (dotted red line).
For all three cases, the distribution was made positive via exponentiation (\ref{expPhi}). Generally, the error is largest for the RTA distribution, and smallest
for the self-consistent $\alpha = 3/2$ exponent 
(except at very early times $\tau \lton 1.3\tau_0$ when the Grad ansatz has 
the largest error). At $\tau = \tau_0$, the power-law Grad form
is completely accurate for any exponent because it admits local
thermal equilibrium distributions. However, the errors quickly grow  at early times,
linearly with time,
because the ansatz cannot describe free streaming exactly.
The overall errors for $K_0 = 1$ are quite large, reaching about $40$\% 
at $\tau \sim 2-3 \tau_0$ but later the accuracy improves rapidly.
For $\tau \gton 13\tau_0$, the self-consistent shear viscous $\delta f$ has
in fact smaller error than the SR ansatz, and for $\tau \gton 19\tau_0$ the
Grad ansatz as well beats the SR form in accuracy.
%

The shaded gray box in Fig.~\ref{Fig:RMS_K1} indicates the inherent error due to
finite statistics in the analysis. 
Because a stochastic Boltzmann solver is employed, 
counts in histogram bins
have random ${\cal O}(1/\sqrt{N})$ relative fluctuations. Therefore, even
a perfect reconstruction model gives $\varepsilon_{RMS} > 0$. The inherent
error for a perfect model can be estimated as
\be
\varepsilon_{MIN} \approx \sqrt{ \frac{1}{N_{bins}}\sum_{ij} \frac{1}{\Delta N_{tr}(p_{T,i-1} \le p_T < p_{T,i}, \xi_{j-1} \le \xi < \xi_j)}} \ \ ,
\ee
which follows because in each bin the fluctuation in the relative error 
between the exact count $\bar N$ and the measured count $N$ can
be estimated as
\be
\left(\frac{\bar N}{N} - 1\right)^2 = \left(\frac{\bar N}{\bar N +\delta N} - 1\right)^2 \approx \frac{\delta N^2}{\bar N^2} \ \to \ \frac{1}{N} \ .
\ee
Note that this is a simple statistical estimate based on the actual 
``measured'' histograms and small $1/\sqrt{N}$ fluctuations;
therefore, the shaded band is not an exact lower bound on $\varepsilon_{RMS}$.

\begin{figure}[h]
\leavevmode
\epsfysize=6.5cm
\epsfbox{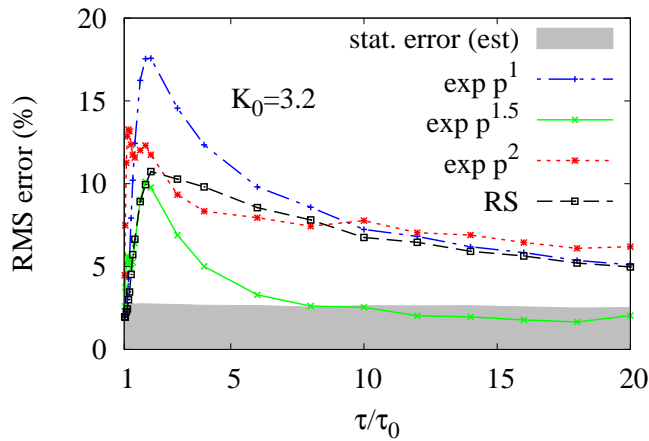}
\caption{The same as Fig.~\ref{Fig:RMS_K1}
but for about three times smaller specific shear viscosity $\eta/s \approx 0.06$ 
(initial inverse Knudsen number $K_0 = 3.2$).}
\label{Fig:RMS_K3.2}
\end{figure}

The comparison changes qualitatively for smaller shear viscosities.
Figure~\ref{Fig:RMS_K3.2} shows the evolution of the overall reconstruction 
error for the four
viscous $\delta f$ models for initial inverse Knudsen number $K_0 = 3.2$ ($\eta/s \sim 0.06$). 
Compared to the $K_0 =1$ case, errors are reduced for all four models.
For the SR ansatz (black solid line), 
the peak error is now only $\approx 10$\%, which gradually
decreases to about $5$\% by late times $\tau \sim 20 \tau_0$. On the other hand,
there is a much bigger separation between the exponentiated 
power-law Grad results for the
different exponents. By and large the worst choice
is still the RTA ansatz ($\alpha = 1$,
dashed-dotted blue line), followed in inaccuracy by the quadratic Grad form 
($\alpha = 2$, dotted red line), while the most accurate reconstruction is 
with the self-consistent BTE corrections ($\alpha = 3/2$, solid green line).
Unlike for $K_0 = 1$, for $K_0 = 3.2$ the Grad ansatz has nearly the same accuracy as the SR form at all times $\tau \gton 2.5\tau_0$, 
and at $\tau \gton 10\tau_0$ the RTA result
also has similar accuracy. The best reconstruction model, however,
is clearly the exponentiated power-law with the self-consistent $3/2$ exponent.
For $K_0 = 3.2$, it has comparable error to the SR ansatz at early $\tau \sim 1.5-2\tau_0$,
and then progressively smaller errors than the SR ansatz and the other two 
power-law
models at all times
$\tau \gton 2\tau_0$. In fact, by $\tau \sim 7\tau_0$ its error becomes as small as
the estimated resolution of our analysis (shaded gray band).
%

It is important to underscore the critical role of exponentiation in these
comparisons. Without exponentiation, RMS errors for the power-law Grad form 
are {\em much} larger. For example, for $K_0 = 1$ ($\eta/s \sim 0.2$) 
the overall error peaks at early times near
 $200$\%, $400$\%, and $600$\% for shear viscous corrections
from the RTA model, the
self-consistent linearized BTE calculation, and the Grad ansatz, respectively.
The same three models still give about $40$\%, $80$\% and $120$\% peak RMS error 
for $K_0 = 3.2$ ($\eta/s \sim 0.06$).

\subsection{Memory effect and improved $\boldsymbol{\delta f}$ models based on time derivative}

The transport solutions depend on both the cross section and initial conditions.
Changing either results, mathematically, in a different solution.
It is then natural to ask to what extent one may hope to accurately reproduce
the phase space density $f$ if only truncated information encoded in the 
local hydrodynamic 
fields is available. 

To get some insight into the sensitivity of the reconstructed phase space
distributions to information not captured in hydrodynamics, i.e., a memory
of the transport evolution, we
compare the full transport solutions at two different but hydrodynamically equivalent
times. Recall the discussion in Sec.~\ref{Sc:fitting}
that the three hydrodynamic quantities available
in the 0+1D Bjorken evolution considered here can be combined into an overall
normalization for $f$, a momentum scale $T_{eff}$, and a dimensionless
shear stress measure $R_\pi$. If one measures $p_T$ relative to $T_{eff}$, and 
takes out the $n \propto 1/\tau$ decrease in the comoving density with time,
then $R_\pi = \pi_L/p$ is the only hydrodynamic quantity that can affect
the rescaled distribution function
\be
\bar f\!\left(\frac{p_T}{T_{eff}}, \xi,\tau\right) \equiv \tau f(p_T, \xi, \tau)  \ .
\ee

\begin{figure}
\leavevmode
\epsfysize=7.5cm
\epsfbox{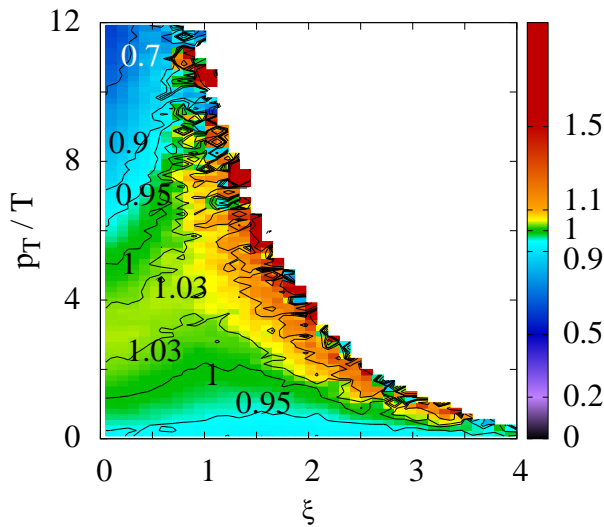}
\caption{Ratio of the rescaled phase space density $\tau f(p_T/T_{eff},\xi,\tau)$
at $\tau = 2\tau_0$ to that at $\tau = 18\tau_0$ as a function of normalized 
transverse momentum $p_T/T_{eff}$ and rapidity $\xi$,
calculated with MPC/Grid for massless particles with 
$\eta/s\sim 0.2$ (initial inverse Knudsen number $K_0 = 1$) and 
isotropic $2\to 2$ scattering. The ratio of counts in rectangular 
bins of size $\Delta (p_T/T_{eff}) \times \Delta \xi \approx 0.16 \times 0.1$ is shown. 
Bins with fewer than 200 counts are not plotted
(empty white region in the plot). 
Contour lines at 0.7, 0.9, 0.95, 1, and 1.03 are also shown.
}
\label{Fig:memory}
\end{figure}

Figure~\ref{Fig:memory} shows the ratio of the rescaled transport solution
at $\tau = 2\tau_0$ to the rescaled solution at $18\tau_0$ as a function of 
$p_T/T_{eff}$ and $\xi$, i.e., 
$\bar f(p_T/T_{eff},\xi,2\tau_0) / \bar f(p_T/T_{eff},\xi,18\tau_0)$,
for the highest shear viscosity $\eta/s \sim 0.2$
we studied (equivalently, the smallest inverse Knudsen number of $K_0 = 1$). 
At both times, $R_\pi \approx -0.168$
(cf. the stars in Fig.~\ref{Fig:Rpi}). 
The white region in the top right half of 
the plot has bins with fewer than 200 counts, therefore, the ratio is not
plotted there. If the phase space densities only depended on hydrodynamic variables,
the ratio would be unity. Instead, there are deviations of up to $\sim 30$\%
in phase space. For example, at the earlier time there are $10-30$\%
fewer high-$p_T$ particles at midrapidity ($p_T \gton 8 T_{eff}$, $|\xi| \lton 1$), and also $\sim 10$\% fewer particles at very low $p_T \lton 0.5 T_{eff}$
 but high rapidity $|\xi| \sim 1-2.5$. The deficit is compensated
by an excess of particles at more intermediate 
$p_T \sim 2-6 T_{eff}$ and $|\xi| \sim 1-2$.
This indicates that for $\eta /s \sim 0.2$,
it is difficult (potentially impossible)
to reconstruct shear viscous phase space densities 
with better than roughly $5-10$\% accuracy in momentum space
from the hydrodynamic fields alone at the given time.
%

Indeed, 
for systems with memory, knowledge of the current snapshot is in general insufficient
to predict the future evolution. A linear theory of memory effects can
be formulated using a memory kernel $G(\tau)$ analogously to
the  text-book treatment of dispersion in classical electrodynamics, in which
the electrical displacement $D$ is given by the field strength $E$ as
\be
D(t) = E(t) + \int_0^\infty d\tau\, G(\tau) E(t - \tau) \ .
\ee
This is an infinite series of time derivatives because Taylor expansion of 
$E(t-\tau)$
leads to
\be
D(t) = E(t) + \sum_{n = 0}^\infty c_n \frac{d^n E(t)}{dt^n}  \ ,
\qquad
c_n \equiv \int_0^\infty d\tau \frac{(-\tau)^n}{n!} G(\tau) \ .
\ee
But if the kernel involves some short time scale $\tau_R$, 
for example, $G(\tau) \sim e^{-\tau/\tau_R}/\tau_R$, then $c_n \sim \tau_R^n$
and one may truncate by only keeping the first derivative $dE/dt$.

\begin{figure}[h]
\leavevmode
\epsfysize=6.5cm
\epsfbox{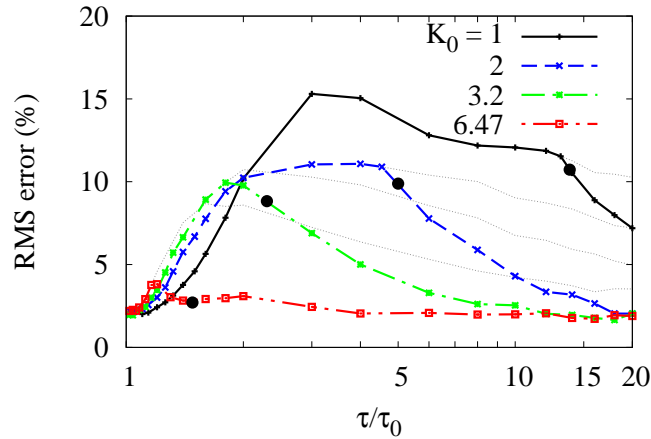}
\caption{Root-mean-squared reconstruction error as a function of normalized
proper time $\tau/\tau_0$ for a shear viscous $\delta f$ model that
switches between the Strickland-Romatschke ansatz and the exponentiated 
power-law Grad form with self-consistent $\alpha = 3/2$ exponent. Results
for initial Knudsen numbers $K_0 = 1$ (thick black line), 2 (dotted blue line),
3.2 (dashed-dotted green line), and 6.47 (dashed-double-dotted red line) are shown.
Error curves for the SR ansatz are also extended to times 
after the switching time (thin dotted
black lines). The filled circles show the switching times for a simple model that
applies the power-law Grad form whenever $R_\pi \equiv \pi_L/p > -0.2$ and 
$dR_\pi/d\tau > 0$, while the SR ansatz otherwise.
}
\label{Fig:switch}
\end{figure}

To demonstrate how first time derivatives
can be used
to improve the modeling of viscous phase space corrections,
we construct a simple model that applies the Strickland-Romatschke 
form during the early departure from equilibrium
but switches to the exponentiated
power-law Grad form with the self-consistent $\alpha = 3/2$ exponent 
when the system relaxes towards local equilibrium at late times.
 Figure~\ref{Fig:switch} shows the overall RMS error
as a function of $\tau/\tau_0$ for $\eta/s \sim 0.2$ 
($K_0 = 1$, thick black line), 0.1 ($K_0 = 2$, dashed blue line), 
0.06 ($K_0 = 3.2$, dashed-dotted green line), and 0.03 ($K_0 = 6.47$, dashed-double-dotted red line). Note
the log scale on the horizontal axis. For all four curves,
the switch is carried out at the optimal proper time $\tau_{switch}$ when
the two models have the same RMS error (thus, $\tau_{switch}$ depends on $K_0$).
To show $\tau_{switch}$ and the higher accuracy gained by switching,
the error curves for the SR ansatz are extended 
in the plot for $\tau \ge \tau_{switch}$ (thin dotted lines).

Of course, a particlization model that requires the full transport solution is not 
very useful. In practice, one can only use information available in
hydrodynamic simulations.
The shear stress to pressure ratio $R_\pi$ is insufficient
because it takes the same value at both early and late times 
(cf. Fig.~\ref{Fig:Rpi}). 
However, if we also use the sign of $dR_\pi/d\tau$,
then we can switch almost optimally with the simple rule below:
\be
\begin{matrix}
 {\rm if}\ R_\pi > -0.2 \ {\rm and} \ \displaystyle\frac{dR_{\pi}}{d\tau} > 0: 
  & {\rm use\ the\ exponentiated\ Grad\ ansatz\ with\ } \alpha=3/2, \cr
{\rm otherwise:}\hfill & {\rm use\ the\ Strickland-Romatschke\ form.}\hfill \cr
\end{matrix}
\ee
The switching times corresponding to the above simple model 
(filled circles in Figs.~\ref{Fig:Rpi}) and \ref{Fig:switch})
close to the most optimal times, except for the largest $K_0 = 6.47$.
In any case, the simple rules above always reduce the RMS 
reconstruction error compared to using the SR form at all times,
and lead to markedly smaller overall error in the reconstructed phase densities
at typical freezeout times $\tau/\tau_0 \sim 5-20$ occuring in 
hydrodynamic simulations of A+A collisions at RHIC and LHC energies.
%

The above results advocate the development of improved viscous phase space
correction models that incorporate not only the local hydrodynamic fields 
but their first time derivatives as well. At present, 
derivatives are not included in the output of
hydrodynamic codes commonly used in heavy-ion physics.
However, printing first derivatives of hydrodynamic quantities should be 
straightforward 
because the hydrodynamic equations of motion are of first order in time.

\section{Conclusions}

In this work we utilized nonlinear $2\to 2$ kinetic theory to assess the
accuracy of shear viscous phase space correction ($\delta f$) models. As in Ref.~\cite{Huovinen:2008te}, a massless one-component gas undergoing 0+1D Bjorken expansion with
specific shear viscosity $\eta/s \approx const$ was studied,
which provides a convenient means to create a system with sustained shear.
We then studied how well four different shear viscous $\delta f$ models: 
the quadratic Grad form (cf. Sec.~\ref{Sc:Grad}), 
the Strickland-Romatschke (SR) ansatz (cf. Sec.~\ref{Sc:SR}),
self-consistent shear viscous corrections from linearized kinetic theory (cf. Sec.~\ref{Sc:lin}), and shear corrections from the relaxation time approach (RTA) 
(cf. Sec.~\ref{Sc:RTA}),
reproduce
the full phase space distribution $f(p_T,\xi,\tau)$ solely from the exact
hydrodynamic fields corresponding to the transport solution.

In general we find that at early times the SR form is the most accurate,
whereas at late times or for small $\eta/s\sim 0.05$ the self-consistent corrections
from kinetic theory perform the best. In addition, we show that 
the positivity of the phase space density in additive $f = f_{\rm eq} + \delta f$ 
shear correction models can be ensured via a simple exponentiation of the correction (cf.
Sec.~\ref{Sc:exp}), which dramatically improves the reconstruction accuracy.

Finally, we demonstrate that there is an inherent uncertainty in $\delta f$ 
reconstruction because the limited information available in the hydrodynamic fields
does not precisely capture the evolution history of the transport (in other words,
the system has memory). We then illustrate 
how even more accurate viscous $\delta f$ models can be
constructed via the incorporation of the first time derivative of hydrodynamic fields,
which are readily available in hydrodynamic simulations (though usually not included
in the output).

Note that this work is limited
to a massless one-component gas undergoing a one-dimensional Bjorken expansion.
It would be very interesting to investigate in the future how well 
shear viscous $\delta f$ 
models for mixtures, such as identified particles in a hadron gas\cite{MolnarWolff},
perform against full kinetic theory. Bulk viscous correction models would
be similarly important to test (this requires nonzero mass). 
Finally, it would be important
to test the improved viscous $\delta f$ models constructed in this work in more 
realistic scenarios for heavy-ion reactions that include both transverse and 
longitudinal expansion and inhomogeneous initial geometry.

\acknowledgements
M.D. and D.M. 
would like to thank the Wigner GPU Laboratory at the Wigner Research Center for Physics (Budapest, Hungary) for providing computing resources and support.
D.M. also thanks the hospitality of the Brookhaven National
Laboratory where parts of this work have been done.
This work was supported by the US Department of Energy under
grant DE-SC0016524, and the Purdue Research Foundation.


\end{document}